%% file: main.tex
\documentclass[11pt]{article}
\pdfoutput=1 

\usepackage[letterpaper, margin=1in]{geometry}
\usepackage[T1]{fontenc}
\usepackage[utf8]{inputenc}
\usepackage{mathptmx}                 
\DeclareFontFamily{OMS}{calcm}{\skewchar\font48}
\DeclareFontShape{OMS}{calcm}{m}{n}{<-6> cmsy5 <6-7> cmsy6 <7-8> cmsy7 <8-9> cmsy8 <9-10> cmsy9 <10-> cmsy10}{}
\DeclareMathAlphabet{\mathcal}{OMS}{calcm}{m}{n}
\usepackage[scaled=0.92]{helvet}      
\usepackage{courier}                  
\usepackage{microtype}

\usepackage{amsmath, amssymb, mathtools}
\DeclareMathOperator*{\argmax}{arg\,max}
\DeclareMathOperator{\SR}{SR}
\newcommand{\E}{\mathbb{E}}
\newcommand{\R}{\mathbb{R}}
\newcommand{\dist}{\mathcal{D}}           
\newcommand{\normal}{\mathcal{N}}
\newcommand{\hist}{\mathrm{hist}}
\newcommand{\future}{\mathrm{future}}

\usepackage{booktabs, multirow, array}
\usepackage{graphicx}
\usepackage[font=small, labelfont=bf, labelsep=period, skip=7pt]{caption}
\usepackage{enumitem}
\usepackage{placeins}
\usepackage{xcolor}
\usepackage{pifont}
\usepackage{tikz}
\usetikzlibrary{arrows.meta}
\usepackage[ruled, linesnumbered, noend]{algorithm2e}
\SetAlFnt{\small}
\SetAlCapFnt{\small}
\SetAlCapNameFnt{\small}

\SetCommentSty{algcomment}
\SetKwInput{KwIn}{Input}
\SetKwInput{KwOut}{Output}

\input{figure_colors}

\usepackage[numbers, sort&compress]{natbib}
\usepackage[colorlinks=true, linkcolor=cBlue, citecolor=cBlue, urlcolor=cBlue,
    pdftitle={Backtesting Trading Strategies with GAN To Avoid Overfitting},
    pdfauthor={Ao Sun, Yuh-Dauh Lyuu}]{hyperref}
\usepackage[capitalize, noabbrev]{cleveref}
\crefname{algocf}{Algorithm}{Algorithms}

\title{\textbf{Backtesting Trading Strategies with GAN To Avoid Overfitting}%
\thanks{This paper is the English version of the first author's master's thesis~\cite{sa2018}, written in Chinese in 2018; some of its statements and claims were already outdated in 2022, when the first English version appeared (arXiv:2209.04895v1). The present edition is a thorough revision of that version: the exposition is rewritten and all figures are redrawn, with the aim of improving readability and of documenting the method and the experimental protocol precisely enough for the results to be reproduced. The experimental design and the conclusions are unchanged.}}
\author{Ao Sun\\[3pt] {\normalsize\texttt{r05922147@ntu.edu.tw}} \and Yuh-Dauh Lyuu\\[3pt] {\normalsize\texttt{lyuu@csie.ntu.edu.tw}}}
\date{{\normalsize Department of Computer Science and Information Engineering, National Taiwan University}}

\begin{document}

\maketitle

\begin{abstract}
Many studies have shown that selecting a trading strategy only because it performs well in sample (IS) invites overfitting: the performance does not carry over out of sample (OOS). Most of them establish that the phenomenon exists without offering a way to avoid it. We propose such a way, built on one premise: a good, that is, non-overfitted, trading strategy should still work well on other paths generated in accordance with the distribution of the historical data. We use a generative adversarial network (GAN) with long short-term memory (LSTM) networks to learn the distribution of the historical time series, and we backtest trading strategies on the paths that the GAN generates, instead of on the single historical path, to avoid overfitting. In experiments on simulated price processes, where the true distribution is known, the GAN learns some properties of geometric Brownian motion. For each combination of price process and trading strategy, we test 100 choices of strategy parameters. GAN-generated paths and Monte Carlo paths from the true process give the same verdict for 73 to 100 of those choices, depending on the process and strategy.
\end{abstract}


\section{Introduction}
\label{sec:introduction}

In algorithmic trading, strategies and backtesting are important elements. A \emph{strategy} is a set of rules that determine when to buy, when to sell, and how much to trade~\cite{carr2014a}. Once a strategy is at hand, the most common practice is to \emph{backtest} it: the strategy is applied to historical data, and its performance there serves as the evaluation of the strategy.

When researchers evaluate many strategy configurations on a single historical price path and select the one with the highest Sharpe ratio, they may select a configuration that performed well largely by chance. Its performance may then deteriorate on a new path. This phenomenon is known as \emph{backtest overfitting}.

\begin{figure}[t]
    \centering
    \includegraphics[width=\linewidth]{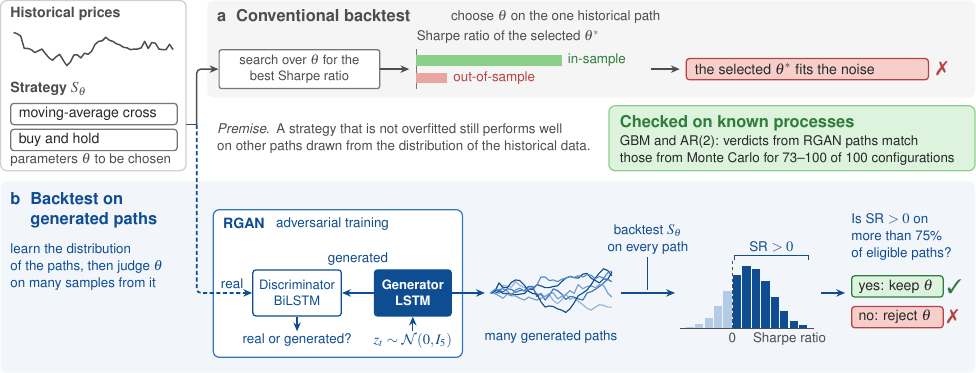}
    \caption{\textbf{Backtesting on generated paths instead of the single historical path.} (a) The conventional backtest searches over the parameters $\theta$ of a strategy $S_\theta$ for the best performance on the historical path; the selected $\theta^*$ fits the noise of that path, and its in-sample Sharpe ratio does not carry over out of sample. (b) Our method trains a recurrent GAN (RGAN) on the historical data (dashed arrow: used for training only), samples many paths from the generator, backtests $S_\theta$ on each of them, and judges $\theta$ by the distribution of the Sharpe ratio (SR), for instance by the share of paths on which it is positive. On simulated processes, for which Monte Carlo paths of the true process are available as a reference, the two sources of paths lead to the same verdict for most configurations (\cref{tab:confusion}).}
    \label{fig:overview}
\end{figure}

The purpose of this study is to propose a backtesting framework based on data generation, in which the data-generating process is learned by a generative adversarial network (GAN). \Cref{fig:overview} summarizes it. Instead of judging a strategy on the one path that history provides, we train a recurrent GAN on the historical data, sample many paths from its generator, backtest the strategy on every one of them, and judge the strategy by the resulting distribution of performance. The premise is that a strategy that is not overfitted should still perform well on other paths drawn from the distribution of the historical data. The study may also extend the use of GANs, which have mostly been applied to images, to financial time series.

We study the framework on simulated price processes, for which the true distribution of paths is known and can serve as a reference. Our contributions are as follows.
\begin{itemize}[leftmargin=*, itemsep=2pt, topsep=4pt]
    \item We state the goal of backtesting as maximizing the expected performance over paths drawn from the data-generating distribution, and show on simulated data how the conventional goal, performance on one path, overfits (\cref{sec:background}).
    \item We learn the distribution of price paths with a recurrent GAN whose generator is an LSTM and whose discriminator is a bidirectional LSTM, and backtest on the paths that it generates (\cref{sec:method}).
    \item We show that the GAN learns some properties of geometric Brownian motion, a commonly used model of stock prices, and study how the number of hidden units, the number of training paths and a scaling of the inputs affect what is learned (\cref{sec:learning}).
    \item We find, for two price processes and two strategies, that the distribution learned by the GAN is sufficient to avoid overfitting to some extent under some of the strategies, and that it can therefore help to screen strategies (\cref{sec:backtest}).
\end{itemize}

\section{Related Work}
\label{sec:related}

There have been many studies of backtest overfitting.

\paragraph{Minimum backtest length.}
\citet{bailey2014a} sharply criticize problems that are routinely overlooked in backtesting, in particular excessive data mining that ignores the growing probability of false positives, and propose the \emph{minimum backtest length} (MinBTL) to control this risk. Their idea is as follows. Draw $N$ numbers independently from the standard normal distribution and let $\E[\max_N]$ be the expected value of their maximum. If $N$ configurations are tried on the same data, a best result as large as $\E[\max_N]$ is to be expected by chance alone and should not be regarded as significant. In terms of the annualized Sharpe ratio, $\E_{\mathrm{annual}}[\max_N] \approx y^{-1/2}\,\E[\max_N]$, where $y$ is the length of the sample in years. Lengthening the sample therefore lowers $\E_{\mathrm{annual}}[\max_N]$ and makes the Sharpe ratio of an experiment more convincing: when, for example, $\E_{\mathrm{annual}}[\max_N]$ is close to zero, a Sharpe ratio above zero is regarded as a significant result instead of a false positive. \citet{bailey2014b} discuss the topic further.

\paragraph{Probability of backtest overfitting.}
\citet{prado2013a} introduce the \emph{probability of backtest overfitting}. Overfitting is defined by
\begin{equation}
    \sum_{n=1}^{N} \E\bigl[\bar{r}_n \mid r \in \Omega_n^*\bigr]\, \Pr\bigl[r \in \Omega_n^*\bigr] < \frac{N}{2},
\end{equation}
where $\E[\bar{r}_n \mid r \in \Omega_n^*]$ is the out-of-sample rank of the strategy that ranks best in sample. The intuition is that overfitting occurs when the strategy with the best in-sample performance performs below average out of sample; more precisely, when the strategy ranked best in sample is expected to rank below the average out of sample. To estimate this, the authors consider a performance matrix $M$ whose entry $M_{st}$ is the performance, such as the profit or loss, of strategy $s$ at time $t$. Given a constant $K$, they divide $M$ into $K$ blocks of equal length along time and combine the blocks into pairs of an in-sample set and an out-of-sample set, on which backtesting and selection are repeated. Note, however, that the method only applies when one strategy has to be chosen among several; its goal is to avoid overfitting in the selection process.

\paragraph{Multiple testing.}
\citet{harvey2015a} treat the problem statistically. When the best configuration (the hyperparameters) of a model is searched for on a single data set and evaluated by the Sharpe ratio, a multiple test has to replace the single test, which means that the Sharpe ratio obtained needs an additional haircut. Although adjusting the Sharpe ratio is common in practice, the authors find that the adjustment is not linear. Generally, a large Sharpe ratio needs a small haircut, because it indicates that the strategy is genuinely profitable; a small Sharpe ratio needs a large one, because it is more likely to be a false positive.

\paragraph{Simulation.}
\citet{carr2014a} discuss how backtest overfitting can be solved from the perspective of simulation. They assume that the price follows an Ornstein--Uhlenbeck (OU) process and estimate the parameters of the process from past data. The fitted model then generates many paths by Monte Carlo simulation, strategies are tested on these paths, and the behaviour of each strategy on them is analysed. Our approach shares this viewpoint, but learns the process that generates the paths with a GAN.

\section{Background: Strategies, Backtests and Overfitting}
\label{sec:background}

This section fixes the definitions used in the paper, shows on simulated data how easily the conventional backtest overfits, and states the goal that our method pursues instead.

\subsection{Strategies}
\label{sec:strategies}

\begin{figure}[t]
    \centering
    \includegraphics[width=\textwidth]{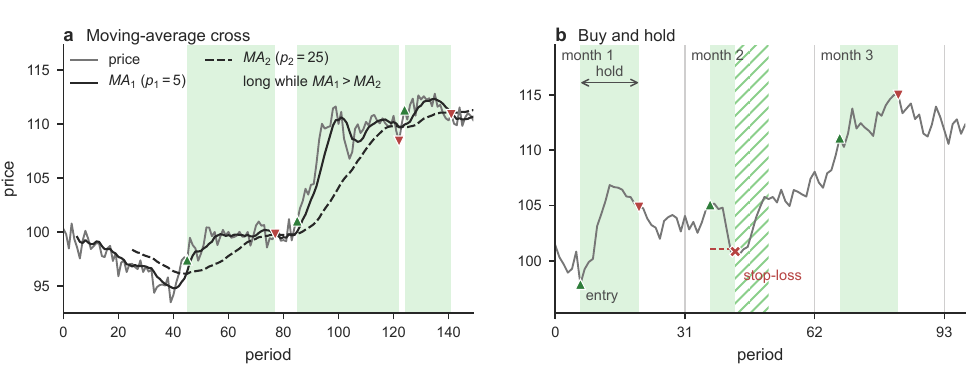}
    \caption{\textbf{The two strategies on a simulated random walk.} (a) MAC with $p_1 = 5$ and $p_2 = 25$ buys ($\blacktriangle$) when the average $\mathit{MA}_1$ rises above $\mathit{MA}_2$ and sells ($\blacktriangledown$) when it falls below; the shaded periods are those with a long position. (b) BH with $\mathit{entry} = 6$, $\mathit{hold} = 14$, $\mathit{stop\text{-}loss} = 4$ and a long side: in every month of 31 periods the position is opened on day 6 and closed 14 days later. In the second month the loss reaches the stop-loss (dashed level) and the position is closed early ($\times$); the hatched periods are the rest of the intended holding period.}
    \label{fig:strategies}
\end{figure}

There seems to be no standard definition of a strategy; the trading rules of \citet{carr2014a}, for instance, describe the actual buying and selling. For the discussion that follows we define a strategy as a function that maps a time series, such as a stock price, to a position at every time step:
\begin{equation}
    S_\theta \colon \{p_t\}_{t=1}^{T} \to \{-1, 0, 1\}^{T},
    \label{eq:strategy}
\end{equation}
where $-1$ stands for a short position, $0$ for no position and $1$ for a long position. A \emph{configuration} of a strategy $S$ is a specific choice of values for its parameters $\theta$. For example, $(p_1,p_2)=(5,25)$ is one configuration of the moving-average strategy described below. We write $S_\theta(p)$ for the positions taken on price path $p$ using those parameter values. We use two strategies, summarized in \cref{tab:strategies} and illustrated in \cref{fig:strategies}.

\paragraph{Moving-average cross (MAC).}
The MAC strategy~\cite{investopedia-a} has two parameters, $p_1$ and $p_2$. At each step, $\mathit{MA}_i$ averages the current price and the preceding $p_i$ prices, for $p_i+1$ observations in all. The strategy takes a long position when $\mathit{MA}_1 > \mathit{MA}_2$ and leaves the market when $\mathit{MA}_1 < \mathit{MA}_2$. When the averages are equal, it keeps its current position.

\paragraph{Buy and hold (BH).}
The BH strategy~\cite{investopedia-b} has four parameters: \emph{entry}, \emph{hold}, \emph{stop-loss} and \emph{side}. It divides the trading days into months and trades once in each month: it enters the market on day \emph{entry}, on the side given by \emph{side}, holds the position for \emph{hold} days and then closes it. If the loss during the holding period exceeds \emph{stop-loss}, the position is closed early.

\begin{table}[t]
    \centering
    \caption{The two strategies and the parameter grids searched in \cref{sec:backtest}. A BH configuration with $\mathit{entry} = 0$ or $\mathit{hold} = 0$ never trades.}
    \label{tab:strategies}
    \small
    \begin{tabular}{@{}llll@{}}
        \toprule
        Strategy & Parameter & Range & Meaning \\
        \midrule
        MAC & $p_1$ & 1--49 & number of preceding prices included in $\mathit{MA}_1$ \\
            & $p_2$ & 1--49 & number of preceding prices included in $\mathit{MA}_2$ \\
        \addlinespace
        BH  & \emph{entry} & 0--29 & day of the month on which the market is entered \\
            & \emph{hold} & 0--20 & number of days the position is held \\
            & \emph{stop-loss} & 0--20 & largest loss tolerated while holding \\
            & \emph{side} & $\{-1, 1\}$ & short or long \\
        \bottomrule
    \end{tabular}
\end{table}

\subsection{Backtesting}
\label{sec:backtesting}

Backtesting simulates a trading strategy on historical data and computes an indicator that helps to judge the strategy. We define a backtest as a function
\begin{equation}
    B \colon \{p_t\}_{t=1}^{T} \times \{-1, 0, 1\}^{T} \to \R
    \label{eq:backtest}
\end{equation}
that maps a path and the positions of a strategy on it to a real number measuring the quality of the strategy. The specific $B$ varies with the trading goal, but we stipulate that a larger value means a better strategy (otherwise $B$ is multiplied by $-1$), and we call the value of $B$ the \emph{performance} of the strategy. Throughout this paper $B$ is the Sharpe ratio,
\begin{equation}
    \SR = \frac{\E[X] - R_f}{\sigma_X},
    \label{eq:sharpe}
\end{equation}
where $X$ is the profit or loss of each trade, that is, of each pair of an entry and an exit that the positions produce on the path, $\sigma_X$ is its standard deviation and $R_f$ is the risk-free return, which we set to zero.

\paragraph{The conventional objective.}
Given a strategy $S$, a trader wants the configuration that maximizes the future performance of the strategy,
\begin{equation}
    \theta^* = \argmax_{\theta}\; B\bigl(p^{\future}, S_\theta(p^{\future})\bigr).
    \label{eq:objective-future}
\end{equation}
Since future data are not available, the trader can only search on past data:
\begin{equation}
    \theta^* = \argmax_{\theta}\; B\bigl(p^{\hist}, S_\theta(p^{\hist})\bigr).
    \label{eq:objective-history}
\end{equation}

\subsection{Backtest overfitting: an example}
\label{sec:overfitting-example}

\begin{figure}[t]
    \centering
    \includegraphics[width=\textwidth]{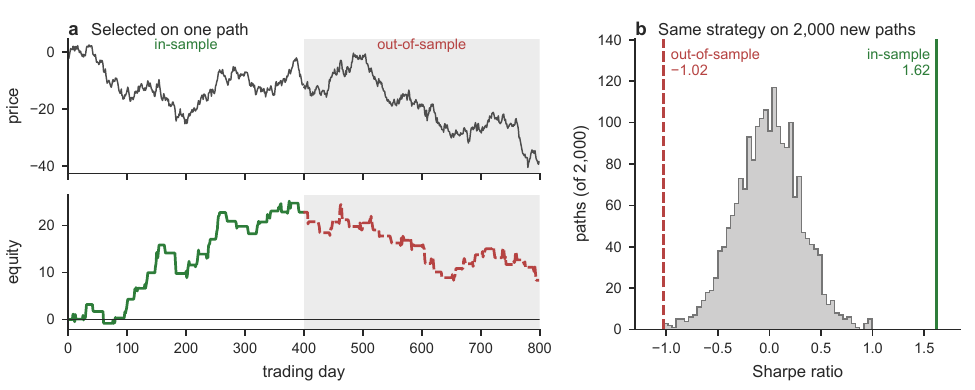}
    \caption{\textbf{A strategy selected in sample on a random walk.} (a) Price path (top) and equity curve (bottom) of the buy-and-hold configuration with the best in-sample Sharpe ratio; the configuration is selected on the first 400 days and applied unchanged to the last 400 days (shaded). (b) Distribution for the same configuration on 2,000 simulated white-noise price paths. The vertical lines mark the annualized Sharpe ratios from panel~a (in sample, $1.62$; out of sample, approximately $-1.02$).}
    \label{fig:overfitting}
\end{figure}

Optimizing \cref{eq:objective-history} on a single price path can lead to backtest overfitting. To see how this happens, consider a buy-and-hold strategy $S'$ applied to a random walk,
\begin{equation}
    P(t+1) = P(t) + \epsilon(t), \qquad \epsilon(t) \sim \normal(0, 1).
    \label{eq:random-walk}
\end{equation}
We simulate 800 days using \cref{eq:random-walk}. Among the configurations of $S'$, we select the one with the highest Sharpe ratio over the first 400 days. We denote this configuration by ${\theta'}^*$ and the resulting strategy by ${S'}^*$. We then apply the same strategy to the remaining 400 days. \Cref{fig:overfitting}a shows both periods on the same price path.

On a random walk, a position chosen using only observations through step $k$ cannot predict the next increment. Let $q_k$ denote the position held over that increment and $\mathcal{F}_k$ the price history through $k$. Over a fixed horizon of $L$ steps the expected trading profit is
\begin{equation}
    \E\!\left[\sum_{k=t}^{t+L-1}q_k\,\epsilon(k)\right] = \sum_{k=t}^{t+L-1}\E\!\left[q_k\,\E[\epsilon(k)\mid\mathcal{F}_k]\right] = 0,
\end{equation}
because each new increment has conditional mean zero. This applies to any configuration fixed before the backtest, including one with a stop-loss: after exit, its position is zero until it enters again.

In \cref{fig:overfitting}a, the upper plot shows the simulated price and the lower plot tracks the equity of ${S'}^*$. The strategy has an annualized Sharpe ratio of $1.62$ during the period used for selection. In the shaded test period, its annualized Sharpe ratio is approximately $-1.02$, as estimated from the equity curve. Here the ratio uses daily trading profits while in the market and is annualized using the number of such observations per year.

The reason is simple. This configuration performs so well in sample because it fits the noise in the sample. For example, the backtest data may happen to rise from the 12th to the 17th of every month, and the selected configuration happens to buy on the 12th and hold for five days. \Cref{eq:random-walk}, however, implies that a rise and a fall between the 12th and the 17th of any later month are equally likely, so this calendar pattern offers no expected profit on a new random walk.

\subsection{Judging a strategy on many paths}
\label{sec:many-paths}

\Cref{fig:overfitting}b illustrates evaluation across many paths. We keep the selected configuration fixed and apply it to 2,000 simulated paths of 500 trading days each, using the entry dates from the selection period. In this panel the prices themselves are independent standard normal draws. In panel~a it was the \emph{price increments} that were independent. The histogram is centered near zero, and none of the 2,000 paths yields a Sharpe ratio as high as the in-sample value of $1.62$. This comparison is specific to the white-noise benchmark. Panel~a provides the out-of-sample comparison under the original random walk.

\begin{figure}[t]
    \centering
    \includegraphics[width=\textwidth]{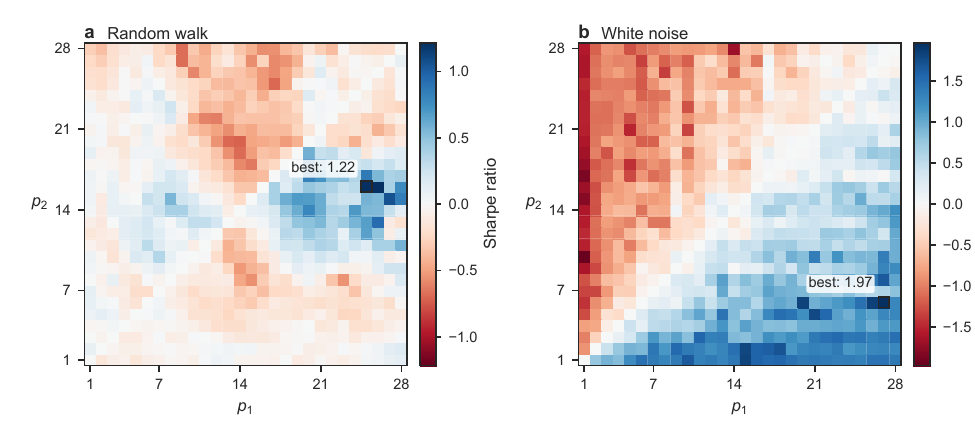}
    \caption{\textbf{Sharpe ratio of MAC for every configuration $(p_1, p_2)$ with $p_1, p_2 \le 28$, on one path.} (a) On a random walk, where no configuration has a positive expected profit, the Sharpe ratio jumps between neighbouring cells, and the best cell (square) rests on four trades. (b) On white noise, where profitable configurations exist, the Sharpe ratio varies smoothly, and the best cell (19 trades) lies in a region of good configurations. Each colour scale is symmetric about zero; configurations that make fewer than three trades, such as those with $p_1 = p_2$, are left at zero.}
    \label{fig:heatmaps}
\end{figure}

Overfitting can also be examined through the smoothness of the performance as a function of the configuration. We use MAC on the random walk of \cref{eq:random-walk} and on white noise,
\begin{equation}
    P(t) = \epsilon(t), \qquad \epsilon(t) \sim \normal(0, 1).
    \label{eq:white-noise}
\end{equation}
Note that MAC has an expected profit of zero on the random walk, whereas on white noise there are configurations with a positive expected profit. \Cref{fig:heatmaps} shows the Sharpe ratio of every configuration of the two parameters on one path of each process: the bluer a cell, the higher the Sharpe ratio and the better the backtest; the redder, the opposite. On the random walk, the dark blue cells, whose Sharpe ratio is clearly above zero, do not form a smooth region but appear in jumps; on white noise the result is much smoother.

In \cref{fig:heatmaps}a, a high Sharpe ratio at one pair of moving-average window lengths can fall sharply at nearby window lengths. Such an isolated peak is a warning sign of overfitting. In \cref{fig:heatmaps}b, performance varies smoothly across neighboring pairs of window lengths, which makes the choice of an exact pair less critical.

\subsection{The objective of this paper}
\label{sec:objective}

The examples show how easily overfitting occurs; to a large extent this is because the $\theta^*$ of \cref{eq:objective-history} fits the noise of the past, especially when the space of configurations is large. Assume that the price path $p$ is produced by a data-generating process with distribution $\dist$. We then replace the goal of the trader by
\begin{equation}
    \theta^* = \argmax_{\theta}\; \E_{p \sim \dist}\Bigl[B\bigl(p, S_\theta(p)\bigr)\Bigr],
    \label{eq:objective-expectation}
\end{equation}
that is, we maximize the expected performance of the strategy on paths sampled from $\dist$. The reason is that taking the expectation may remove the noise of any particular path. The idea is similar to bagging in machine learning~\cite{breiman1996a}: we let the paths vote on whether the strategy is good or bad. In practice the expectation is approximated by the mean over a sample of paths,
\begin{equation}
    \theta^* = \argmax_{\theta}\; \frac{1}{N} \sum_{i=1}^{N} B\bigl(p^{(i)}, S_\theta(p^{(i)})\bigr), \qquad p^{(1)}, \dots, p^{(N)} \sim \dist.
    \label{eq:objective-sample}
\end{equation}
A good configuration is thus one that maximizes the mean performance over paths sampled from the data-generating process. The sample of paths gives more than this mean: it gives the whole distribution of the performance, which \cref{sec:backtest} uses to judge a configuration. For real prices $\dist$ is unknown; the next section learns it from data.

\section{Method: Learning the Distribution of Paths with an RGAN}
\label{sec:method}

To evaluate \cref{eq:objective-sample} we need to sample paths from $\dist$. We learn a sampler from the historical data with a GAN built from recurrent networks, and backtest on the paths that it generates (\cref{fig:overview}b).

\subsection{LSTM, GAN and RGAN}
\label{sec:rgan-background}

The long short-term memory (LSTM) network was first proposed by \citet{hochreiter1997a}. Unlike a feed-forward network, an LSTM uses not only the current input but also the results computed at the previous time steps, which makes it suited to sequences.

Generative adversarial networks, proposed by \citet{goodfellow2014a} in 2014, have sparked a wave of research in recent years, with countless variants. Almost all of them share one idea. A generator $G$ maps a distribution $Z$ on a latent space to the target distribution: a sample is drawn from the latent space and passed to the generator, which turns it into a sample of the target distribution. A discriminator $D$ is trained at the same time to tell whether a sample comes from the target distribution or has been produced by the generator from $Z$, and the generator is trained to fool it.

The recurrent GAN (RGAN) was proposed by \citet{hyland2017a} in 2017; the R indicates that both the generator and the discriminator are recurrent networks. Its purpose was to address the privacy problem that arises when the data of real patients are used for research. The data there are time series with four dimensions: oxygen saturation measured by pulse oximetry, heart rate, respiratory rate and mean arterial pressure. If we do not care about the physical meaning of the values, we can equally well take the opening, highest, lowest and closing prices of a stock as the data and bring them into the RGAN framework. In addition, we replace the LSTM of the RGAN discriminator by a bidirectional LSTM, in order to strengthen the discriminator.

\subsection{Model}
\label{sec:model}

\Cref{fig:architecture} gives an overview of our model. At every time step $t = 1, \dots, T$ a vector $z_t$ is sampled from the latent space and passed to the generator. The output of the generator, $\hat{y} = (\hat{y}_1, \dots, \hat{y}_T)$, has the same shape as a real path $y = (y_1, \dots, y_T)$, so either can be passed to the discriminator. Because both are time series, the discriminator produces an output at every time step. Once all time steps have been processed, the outputs of the discriminator for the generated data and for the real data are collected, and the cross-entropy over them is the loss.

\begin{figure}[t]
    \centering
    \makebox[\textwidth][c]{\input{fig_architecture}}
    \caption{\textbf{The RGAN, unrolled over the $T$ steps of a path.} The generator turns latent noise $z_1, \dots, z_T$ into a path $\hat{y}_1, \dots, \hat{y}_T$. The discriminator reads a path, generated (blue) or real (grey), in both directions and outputs at every step the probability $d_t$ that the path is real. Generated paths are used in the same way as real ones in a backtest.}
    \label{fig:architecture}
\end{figure}

\paragraph{Latent space.} The latent space has five dimensions, each an independent standard normal distribution: $z_t \sim \normal(0, I_5)$, independently over $t$.

\paragraph{Generator.} The generator is an LSTM with $u$ hidden units and the sigmoid activation in its recurrent layer, followed at every time step by an output layer with the tanh activation. \Cref{sec:learning} compares several values of $u$.

\paragraph{Discriminator.} The discriminator is a recurrent network with $u$ hidden units per direction, again with the sigmoid activation in its recurrent layer; we tried a unidirectional and a bidirectional LSTM and use the bidirectional one. At every time step an output layer maps the hidden states of both directions to $d_t \in (0, 1)$, the probability that the path is real.

\paragraph{Loss.} With $d_t(\cdot)$ the output of the discriminator at step $t$, the discriminator is trained to minimize the binary cross-entropy
\begin{equation}
    \mathcal{L}_D = -\frac{1}{T} \sum_{t=1}^{T} \Bigl[\log d_t(y) + \log\bigl(1 - d_t(G(z))\bigr)\Bigr],
\end{equation}
averaged over a batch of real paths $y$ and of generated paths $G(z)$, and the generator is trained to make the discriminator label its paths as real, $\mathcal{L}_G = -\frac{1}{T} \sum_{t=1}^{T} \log d_t(G(z))$. The two networks are updated in turn, once per batch.

\paragraph{Size.} The part of the model with the greatest impact is the number $u$ of hidden units of the LSTMs; \cref{tab:parameters} shows how it determines the number of parameters.

\begin{table}[t]
    \centering
    \caption{Number of parameters as a function of the number of hidden units $u$, for a five-dimensional latent space and a one-dimensional path. The counts are those of the implementation, $4u(u+5) + 9u + 1$ for the generator and $8u(u+1) + 20u$ for the discriminator; they include two bias vectors per LSTM gate and $4u$ parameters in the output layer of the discriminator, which applies one weight vector of size $2u$ with $2u$ different biases and averages the loss over the resulting $2u$ outputs per step.}
    \label{tab:parameters}
    \small
    \begin{tabular}{@{}rrrr@{}}
        \toprule
        Hidden units $u$ & Generator & Discriminator & Total \\
        \midrule
        100 & 42,901 & 82,800 & 125,701 \\
        50  & 11,451 & 21,400 & 32,851 \\
        10  & 691    & 1,080  & 1,771 \\
        \bottomrule
    \end{tabular}
\end{table}

\subsection{Preprocessing and scaling}
\label{sec:preprocessing}

Because the last activation of the generator is tanh, its outputs lie in $(-1, 1)$, and the samples have to be normalized to this interval. We use
\begin{equation}
    x_{\mathrm{new}} = 2\,\frac{x_{\mathrm{raw}} - x_{\min}}{x_{\max} - x_{\min}} - 1,
    \label{eq:preprocessing}
\end{equation}
where $x_{\mathrm{raw}}$ is a value of a training path, $x_{\min}$ and $x_{\max}$ are the smallest and the largest value in the whole training set, and $x_{\mathrm{new}}$ is the value that the RGAN is trained on.

\Cref{eq:preprocessing} maps the largest and the smallest value of the training set to $1$ and $-1$. A recurrent generator can be run for more than the $T$ steps it was trained on. When the RGAN generates a time series longer than $T$, however, this normalization causes a problem: the largest value that can be generated is limited. We therefore change the preprocessing to
\begin{equation}
    x_{\mathrm{new}} = \frac{1}{\lambda}\Bigl(2\,\frac{x_{\mathrm{raw}} - x_{\min}}{x_{\max} - x_{\min}} - 1\Bigr),
    \label{eq:scaling}
\end{equation}
where the \emph{scaling} $\lambda \ge 1$ is a constant of our choosing; $\lambda = 1$ recovers \cref{eq:preprocessing}. Generated values are mapped back to prices by inverting \cref{eq:scaling}. \Cref{sec:scaling} examines the effect of $\lambda$.

\section{Experiment I: Learning a Known Process}
\label{sec:learning}

We use geometric Brownian motion (GBM), a common model of stock prices, to examine which properties of the price process the GAN learns and how the size of the training set affects the fit.

\subsection{Setup}
\label{sec:learning-setup}

\paragraph{Process.} The target process of the RGAN is GBM,
\begin{equation}
    dy_t = \mu\, y_t\, dt + \sigma\, y_t\, dW_t,
    \label{eq:gbm}
\end{equation}
where $W_t$ is a Brownian motion. Solving \cref{eq:gbm} gives
\begin{equation}
    y_t = y_0 \exp\Bigl(\bigl(\mu - \tfrac{\sigma^2}{2}\bigr)\, t + \sigma W_t\Bigr),
    \label{eq:gbm-solution}
\end{equation}
so that, for $y_0 = 1$,
\begin{equation}
    \ln y_t \sim \normal\Bigl(\bigl(\mu - \tfrac{\sigma^2}{2}\bigr)\, t,\; \sigma^2 t\Bigr).
    \label{eq:gbm-law}
\end{equation}
Here $t$ is time. We use $\mu = 0.1$, $\sigma = 0.1$ and $y_0 = 1$, and observe a path at equally spaced steps of size $\Delta t = 0.01$; with a slight abuse of notation, $y_t$ also denotes the value at step $t$, whose time is $t\,\Delta t$.

We choose this process for the following reason. Evaluating a GAN, and testing whether a time series comes from the same distribution as a given sample, are open questions~\cite{hyland2017a, theis2015a}. In the field of images, where GANs are most common, the criterion is mostly the human eye, and it is hard to apply the same standard to time series. \Cref{eq:gbm-law}, in contrast, allows a statistical test. Given a collection of generated time series $\hat{y}$, we can examine the distribution of $\ln \hat{y}_t$ across the collection:

\begin{quote}
    \textbf{Hypothesis 1.} For every step $t$, $\ln \hat{y}_t$ follows a normal distribution with mean $\bigl(\mu - \sigma^2/2\bigr)\, t\,\Delta t$ and variance $\sigma^2\, t\,\Delta t$.
\end{quote}
We examine two implications of this hypothesis: the mean and the variance of $\ln \hat{y}_t$ should each grow linearly with elapsed time (\cref{sec:learning-evaluation}).

\paragraph{Training data.} The training set consists of $N$ paths of the same length $T$ simulated from the model; we fix $T = 100$ steps. The number of paths $N$ matters. Here the samples come from a given model, but for real prices there is no such model, and the amount of data is smaller than one might think: to obtain samples, the historical series has to be cut into pieces under some assumption. If, say, the prices of different months are assumed to be statistically similar, ten years of data of one stock yield only 120 samples. We therefore study how $N$ affects the result (\cref{sec:units-samples}). The paths are normalized by \cref{eq:scaling}, which reduces to \cref{eq:preprocessing} for $\lambda = 1$.

\paragraph{Training.} The RGAN of \cref{sec:model} is trained with a batch size of 50. The discriminator is updated by stochastic gradient descent with a learning rate of 0.1 and the generator by Adam with its default settings.

\subsection{Evaluation}
\label{sec:learning-evaluation}

Different models call for different evaluations, and testing the distribution of time series is hard~\cite{hyland2017a, theis2015a}. Because we chose GBM, however, we can take the natural logarithm of the value at every step: by \cref{eq:gbm-law} the distribution at every step is then normal, and its mean and its variance are linear in time.

The criterion for judging during training whether a GAN is learning well is equally vague~\cite{hyland2017a}. Since the time series that we chose has such a distinctive feature, we use the theoretical linear relations as the target: we generate paths with the current generator, compute the mean and the variance of their logarithms at every step, and measure by the coefficient of determination $R^2$ how well they agree with the target over the $T$ steps. In the implementation the target values are the per-step mean and variance of the training paths, which follow the two lines up to sampling error. When both $R^2$ values are close to one, as for the model shown in \cref{fig:gbm}, training can be stopped.

\subsection{Statistical properties of the generated paths}
\label{sec:learning-results}

\begin{figure}[t]
    \centering
    \includegraphics[width=\textwidth]{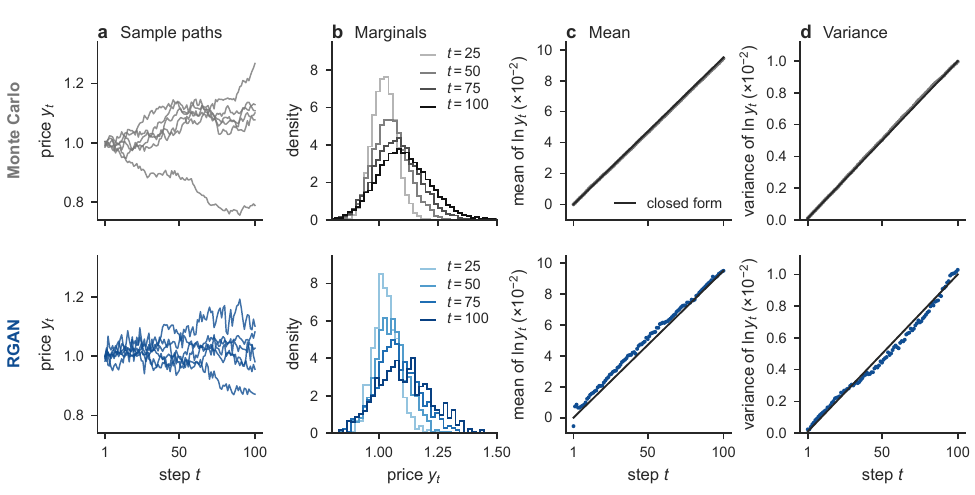}
    \caption{\textbf{Paths of geometric Brownian motion: Monte Carlo (top) against the trained RGAN (bottom).} (a) Six sample paths. (b) Distribution of the price at four steps, from 10,000 simulated and 800 generated paths. (c, d) Mean and variance of $\ln y_t$ at every step $t$ (dots), with the values implied by the GBM (line). The RGAN was trained on $N = 10{,}000$ paths with $u = 100$ and $\lambda = 1$.}
    \label{fig:gbm}
\end{figure}

The top row of \cref{fig:gbm} gives the Monte Carlo benchmark. Panel~b shows the marginal distribution of the price, which is lognormal under GBM. As time increases, the distribution shifts to the right and spreads out. Panels~c and~d show the mean and variance of the \emph{log price}, the quantities for which \cref{eq:gbm-law} predicts linear growth.

The bottom row shows the same quantities for the trained RGAN. The generated price distributions also shift to the right and spread out (panel~b). Their log-price means and variances track the benchmark curves over the training horizon (panels~c and~d). At the retained checkpoint, the monitor of \cref{sec:learning-evaluation} gives $R^2 = 0.994935$ for the mean and $R^2 = 0.9946425$ for the variance.

\subsection{Hidden units and number of training paths}
\label{sec:units-samples}

The previous section shows that, by these measures, the RGAN learns GBM quite well. Training an RGAN is not easy, however: it can stall in a local optimum, and the result depends rather strongly on the choice of hyperparameters such as the learning rate~\cite{bergstra2012a}. In this section we examine how the result depends on the model, through the number of hidden units $u$, and on the data, through the number of training paths $N$, the importance of which was discussed in \cref{sec:learning-setup}.

\begin{figure}[t]
    \centering
    \includegraphics[width=\textwidth]{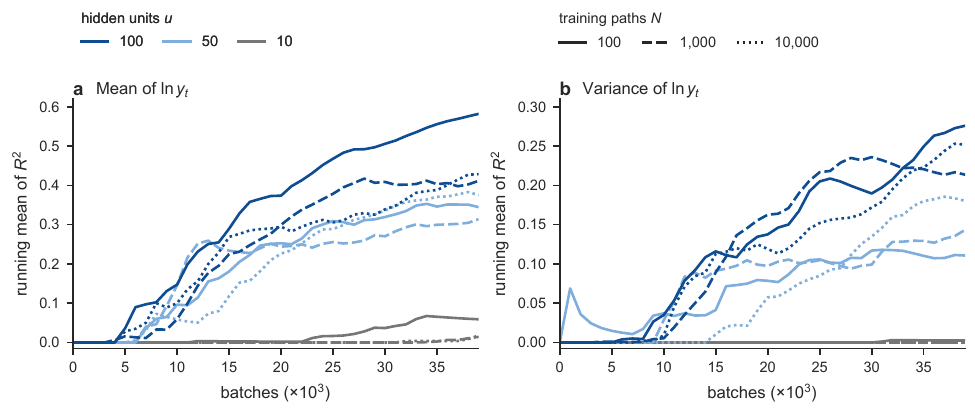}
    \caption{\textbf{Learning GBM with different numbers of hidden units $u$ and of training paths $N$.} Running mean, over the training so far, of the $R^2$ between the generated and the reference per-step mean (a) and variance (b) of $\ln y_t$; the monitor is read every 100 batches, a negative $R^2$ is counted as zero, and $\lambda = 1$. Models with 10 hidden units barely learn, whereas with 100 hidden units the run trained on 100 paths compares favorably with the run trained on 10,000.}
    \label{fig:training}
\end{figure}

\Cref{fig:training} compares training curves across network sizes and numbers of training paths. The horizontal axis is the number of batches fed to the model, with a fixed batch size of 50; the vertical axis is the $R^2$ of \cref{sec:learning-evaluation} for the mean (panel~a) and for the variance (panel~b). Because the monitor fluctuates strongly from one reading to the next, the figure shows its running mean up to that batch, which stays well below the best values that a model reaches.

At each training-set size tested, the networks with 10 hidden units have lower running scores than the networks with 50 or 100 units. For a broader discussion of model size and generalization, see \citet{kawaguchi2017a}.

With $u=100$, the run trained on 100 paths attains running scores comparable to those of the run trained on 10,000 paths for both moments (\cref{fig:training}). These scores compare generated paths with training-set moments. Memorizing training paths could also produce a good fit on this measure.

\subsection{Scaling}
\label{sec:scaling}

\begin{figure}[!tb]
    \centering
    \includegraphics[width=\textwidth]{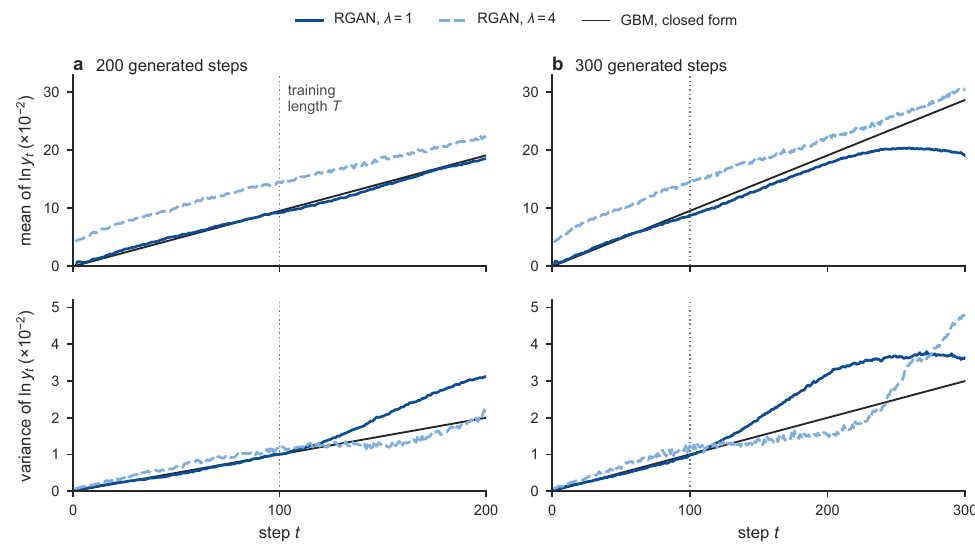}
    \caption{\textbf{Generating beyond the training length $T = 100$, with scaling $\lambda = 1$ and $\lambda = 4$.} Mean (top) and variance (bottom) of $\ln \hat{y}_t$ over 200 (a) and 300 (b) generated steps, for the RGANs trained on $N = 10{,}000$ paths with $u = 100$; the curves are those of the thesis. With $\lambda = 1$ the mean keeps rising until about 250 steps, increasingly below the GBM line, and then decreases, while the variance rises above the line soon after $T$. With $\lambda = 4$ the mean stays linear over all 300 steps, shifted upwards from the first step on; its variance stays below that of $\lambda = 1$ until about 250 steps and rises sharply afterwards.}
    \label{fig:scaling}
\end{figure}

\Cref{sec:preprocessing} introduced the scaling $\lambda$ to let the generator produce paths longer than the $T = 100$ steps it was trained on. \Cref{fig:scaling} shows the mean and the variance of $\ln \hat{y}_t$ when the trained generator is unrolled for 200 and for 300 steps.

Without scaling ($\lambda = 1$) and with 200 generated steps, the mean remains stable, in that it keeps growing linearly with the same slope, whereas the variance shows a clear turn once $t$ exceeds $T$. With 300 generated steps the variance does not remain linear either, and the mean even decreases. The likely reason is the normalization of \cref{eq:preprocessing}, which takes the largest value of the sample as the largest value of the population: the last activation of the generator is tanh, so its range is fixed to $(-1, 1)$, with $1$ and $-1$ corresponding to the largest and the smallest value of the training set, so this is what happens when the GAN tries to generate larger values.

With $\lambda = 4$, the values $0.25$ and $-0.25$ correspond to the largest and the smallest value of the training set, and the problem is alleviated to some extent: the mean keeps growing linearly and smoothly, in particular beyond $T = 100$. There is a clear cost, however: the variance of the generated samples is compressed, which to some extent changes the statistical properties of what is learned.

Within the training horizon the RGAN can reproduce the first two log-price moments of GBM. In the runs compared here, increasing network size improved the training scores more consistently than increasing the number of paths. Beyond the training horizon, scaling extends the range of generated prices but also changes their variance. We next ask how these properties affect the evaluation of trading strategies.

\FloatBarrier
\section{Experiment II: Backtesting on Generated Paths}
\label{sec:backtest}

We now look at the GAN from the perspective of trading strategies. In practice GANs are hard to train~\cite{bojanowski2017a}, and what a GAN learns is sometimes not ideal. Is it then of no help at all for backtesting?

Consider a strategy on a random walk. Its expected one-step profit is zero whenever the next price increment has mean zero conditional on the past. A generator can preserve that expectation even if it misses other features of the increments. Other quantities, including the distribution of the Sharpe ratio, may still differ. This motivates evaluating generated paths with the trading strategies for which they will be used.

Our approach in this section is to use assumed price processes and to study, on these prices, strategies that can theoretically have a positive expected profit as well as strategies whose expected profit is zero. This gives a reference for the out-of-sample performance of a strategy, against which different generation methods can be compared, to see whose results lie closer to the reference.

\subsection{Processes and expected outcomes}
\label{sec:backtest-setup}

In addition to GBM we use the autoregressive process AR(2),
\begin{equation}
    y_t = a + b\, y_{t-1} + c\, y_{t-2} + \epsilon_t, \qquad \epsilon_t \sim \normal(0, 1),
    \label{eq:ar2}
\end{equation}
with $a = 0$, $b = 1.1$ and $c = -0.5$. We choose these values as a contrast to GBM: with them the roots of the characteristic polynomial of \cref{eq:ar2} have an absolute value below one, so the recursion is stable. The simulated paths start from two zeros and approach the stationary distribution. The process takes negative values, which does not matter here: both strategies trade on price differences only. The following arguments motivate the expected profit benchmarks for each strategy and process; \cref{tab:theory} summarizes the four cases.

\begin{description}[leftmargin=1.2em, itemsep=1pt, topsep=4pt]
    \item[BH on GBM.] With positive drift $\mu$, a long position held for a fixed time $h>0$ has positive expected profit because $\E[y_{t+h}\mid y_t]=y_t e^{\mu h}$.
    \item[BH on AR(2).] The mean price is zero at every step under the chosen initialization. With entry and exit fixed by the calendar, the expected price change is therefore zero. This gives the fixed-horizon benchmark. A stop-loss makes the exit time depend on the price path.
    \item[MAC on GBM.] Under GBM, future log returns are independent of past prices. With positive drift, the expected price change while MAC holds a long position is positive.
    \item[MAC on AR(2).] The benchmark is mean reversion. With a long lookback $p_1$ and a short lookback $p_2$, MAC takes a long position when the short average falls below the long one. We therefore expect some configurations to profit from subsequent price recoveries for the coefficients used here.
\end{description}

\begin{table}[t]
    \centering
    \caption{Expected profit for the two strategies under each price process. The BH benchmark uses fixed entry and exit dates. The arguments in \cref{sec:backtest-setup} give these theoretical expectations. \Cref{tab:confusion} compares the empirical verdicts from the two sources of paths.}
    \label{tab:theory}
    \small
    \begin{tabular}{@{}lll@{}}
        \toprule
        & \multicolumn{2}{c}{Process} \\
        \cmidrule(l){2-3}
        Strategy & GBM (upward drift) & AR(2) (mean-reverting) \\
        \midrule
        BH  & yes, on the long side & zero for a fixed holding period \\
        MAC & yes, long positions earn the drift & yes, buy below the mean and sell above \\
        \bottomrule
    \end{tabular}
\end{table}

For each process we train an RGAN with the architecture of \cref{sec:model} ($u = 100$, $N = 10{,}000$ training paths of $T = 100$ steps, $\lambda = 1$).

\subsection{Evaluation}
\label{sec:backtest-evaluation}

As noted by \citet{hyland2017a}, evaluating the quality of data generated by a GAN is challenging. Unlike for traditional machine-learning models, the losses of the generator and of the discriminator say little about the performance of the model, and for images the assessment mostly relies on human ratings. \citet{hyland2017a} therefore propose TSTR (train on synthetic, test on real) and TRTS (train on real, test on synthetic): a traditional classifier is trained on the generated (real) data set and tested on the real (generated) data, and if the two results are close, the GAN is considered to have learned.

We adapt this idea to backtesting (\cref{alg:evaluation}). We train a generator on the training set, find the best-performing strategy on the data that it generates, and record the score of this strategy, such as its Sharpe ratio, on the generated data. Finally we apply the strategy to the test set, obtain its score there, and compare the two scores.

\begin{algorithm}[t]
    \caption{Evaluating an RGAN by backtesting.}
    \label{alg:evaluation}
    \KwIn{data $\mathcal{Y}$; a strategy $S_\theta$ with configurations $\Theta$; a backtest $B$, where $B(\mathcal{Y}', S)$ is the mean performance of $S$ over the paths of $\mathcal{Y}'$}
    \KwOut{a score that tells how well backtests on generated paths carry over to the test data}
    $\mathcal{Y}_{\mathrm{train}}, \mathcal{Y}_{\mathrm{test}} \gets \mathrm{split}(\mathcal{Y})$\;
    $G, D \gets \mathrm{train\_GAN}(\mathcal{Y}_{\mathrm{train}})$\;
    $\mathcal{Y}_{\mathrm{syn}} \gets$ paths sampled from $G$\;
    $\theta^* \gets \argmax_{\theta \in \Theta} B(\mathcal{Y}_{\mathrm{syn}}, S_\theta)$ \tcp*{best configuration on the generated paths}
    $s_{\mathrm{syn}} \gets B(\mathcal{Y}_{\mathrm{syn}}, S_{\theta^*})$\;
    $s_{\mathrm{test}} \gets B(\mathcal{Y}_{\mathrm{test}}, S_{\theta^*})$\;
    \KwRet{$\mathrm{score}(s_{\mathrm{syn}}, s_{\mathrm{test}})$} \tcp*{how close the two scores are}
\end{algorithm}

\Cref{alg:evaluation} describes evaluation when configurations are selected on synthetic data. In the controlled experiments below, the price process is known, so we can compare the two sources directly for the same configuration. For \cref{fig:backtest} we select a configuration on one simulated history and compare its two Sharpe ratio distributions. For \cref{tab:confusion} we sample configurations from the grid and compare their verdicts.

\subsection{Distribution of the Sharpe ratio}
\label{sec:backtest-distributions}

\begin{figure}[t]
    \centering
    \includegraphics[width=\textwidth]{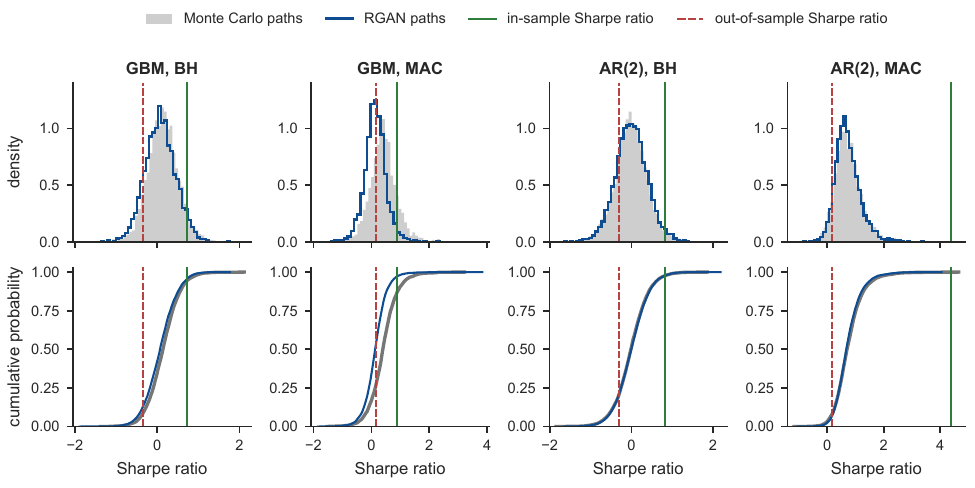}
    \caption{\textbf{Sharpe ratio of the configuration selected in sample, on 5,000 Monte Carlo paths of the true process and on 5,000 RGAN paths.} Density (top) and cumulative distribution (bottom) for each process and strategy; the vertical lines mark the Sharpe ratio of the configuration on the in-sample and on the out-of-sample half of the path used for the selection. Paths with fewer than five trades are left out, which affects MAC only (GBM: 2,970 Monte Carlo and 4,549 RGAN paths remain; AR(2): 4,884 and 4,946).}
    \label{fig:backtest}
\end{figure}

As shown in \cref{sec:learning}, a GAN can approximate common price processes. Is the approximation good enough for backtesting? For each process we simulate a path of 600 steps, select on its first half the configuration on the grid of \cref{tab:strategies} with the best in-sample Sharpe ratio, and compute the Sharpe ratio of this configuration on 5,000 paths of 300 steps drawn from the true process and on 5,000 paths of 300 steps generated by the RGAN, which is thus run beyond the 100 steps it was trained on. \Cref{fig:backtest} shows the two distributions for the four combinations of a process and a strategy.

For BH on GBM the two densities are quite close. In this combination, the model learned by the GAN is thus about as informative a reference for backtesting as the true model.

For MAC on GBM the Monte Carlo distribution has a higher mean than the RGAN distribution. A shift of this kind can change the fraction of paths with positive Sharpe ratios. \Cref{sec:backtest-confusion} examines how often the verdicts agree across configurations.

The two combinations on AR(2) perform well, and the conclusions are similar to those for BH on GBM.

\subsection{Agreement of the verdicts}
\label{sec:backtest-confusion}

For each process and strategy, we sample 100 configurations without replacement from the grid in \cref{tab:strategies}. All configurations share the same 100 Monte Carlo paths and the same 100 RGAN paths. We assess each source separately. For a given configuration, we retain paths with at least five trade outcomes and a defined Sharpe ratio. We call the configuration \emph{effective} if more than 75\% of its retained paths have $\SR>0$. If no paths qualify, the configuration is classified as ineffective. This criterion measures frequency: positive expected profit alone does not ensure that a configuration meets the threshold. Each row of \cref{tab:confusion} counts configurations by the two verdicts.

\begin{table}[!htb]
    \centering
    \caption{Verdicts on 100 configurations per process and strategy. The four middle columns report which sources classify a configuration as effective. Agreement counts both-positive and both-negative verdicts. Expected profit summarizes the benchmarks in \cref{tab:theory}, using fixed holding periods for BH.}
    \label{tab:confusion}
    \small
    \begin{tabular}{@{}lllccccc@{}}
        \toprule
        & & & \multicolumn{4}{c}{Judged effective by} & \\
        \cmidrule(lr){4-7}
        Process & Strategy & Expected profit & both & RGAN only & MC only & neither & Agreement \\
        \midrule
        GBM   & BH  & positive for some & 6  & 2 & 22 & 70  & \textbf{76}  \\
        GBM   & MAC & positive for some & 32 & 9 & 18 & 41  & \textbf{73}  \\
        AR(2) & BH  & zero & 0  & 0 & 0  & 100 & \textbf{100} \\
        AR(2) & MAC & positive for some & 39 & 1 & 1  & 59  & \textbf{98}  \\
        \bottomrule
    \end{tabular}
\end{table}

For BH on AR(2), both sources classify all 100 sampled configurations as ineffective. For MAC on AR(2), they agree on 98 of the 100 configurations.

On GBM, the two sources agree on 76 of the 100 BH configurations. Of these, both classify six as effective and 70 as ineffective. Among the 24 disagreements, 22 configurations meet the threshold on Monte Carlo paths alone and two on RGAN paths alone. For MAC, the two sources agree on 73 configurations. The lower agreement for MAC may reflect the difference between its Sharpe ratio distributions in \cref{fig:backtest}.

Across the four comparisons, agreement with the Monte Carlo reference is highest on AR(2). The GBM results show why the direction of disagreement also matters when using generated paths to screen strategies: for BH, the RGAN rejects most of the configurations that meet the Monte Carlo threshold.

\section{Conclusion}
\label{sec:conclusion}

We investigated whether a GAN can learn a distribution of stock prices that is useful for screening trading strategies. We examined how network size and the number of training samples affect the fit, then compared strategy evaluations on generated paths with evaluations on paths drawn from the known price process.

For GBM, the trained generator reproduced the growth in the mean and variance of log prices over the training horizon. Across the tested network sizes and training sets, the moment scores depended more consistently on the number of hidden units than on the number of training paths. Scaling expanded the attainable price range beyond the training horizon but altered the generated variance. In the strategy experiments, RGAN and Monte Carlo paths gave the same verdict for 73 to 100 of the 100 configurations in each comparison. These results support using generated paths to screen strategies for sensitivity to a single historical path, with the closest agreement observed on AR(2).

\bibliographystyle{unsrtnat}
\bibliography{references}

\end{document}

%% file: figure_colors.tex
\definecolor{cBlue}{HTML}{0F4D92}
\definecolor{cBlue2}{HTML}{3775BA}
\definecolor{cBlueL}{HTML}{DCE8F5}
\definecolor{cGreen}{HTML}{2E7D3A}
\definecolor{cGreenM}{HTML}{8BCF8B}
\definecolor{cGreenL}{HTML}{DDF3DE}
\definecolor{cRed}{HTML}{B64342}
\definecolor{cRedM}{HTML}{E9A6A1}
\definecolor{cRedL}{HTML}{F6CFCB}
\definecolor{cGrey}{HTML}{CFCECE}
\definecolor{cGreyL}{HTML}{ECECEC}
\definecolor{cGreyM}{HTML}{767676}
\definecolor{cGreyD}{HTML}{4D4D4D}
\definecolor{cInk}{HTML}{272727}
\definecolor{cPanel}{HTML}{F5F5F5}

%% file: fig_architecture.tex
%
\begin{tikzpicture}[x=1cm, y=1cm,
    font=\sffamily\fontsize{7.4}{8.8}\selectfont, text=cInk,
    >={Stealth[length=1.5mm, width=1.25mm]},
    gcell/.style={draw=cBlue, fill=cBlue, text=white, rounded corners=1.5pt, line width=0.6pt,
        minimum width=1.5cm, minimum height=0.56cm, inner sep=0pt,
        font=\sffamily\bfseries\fontsize{6.8}{7.8}\selectfont},
    head/.style={draw=cBlue, fill=cBlueL, text=cBlue, rounded corners=1.2pt, line width=0.6pt,
        minimum width=1.5cm, minimum height=0.36cm, inner sep=0pt,
        font=\sffamily\fontsize{6.8}{7.6}\selectfont},
    dcell/.style={draw=cBlue, fill=white, text=cBlue, rounded corners=1.5pt, line width=0.6pt,
        minimum width=1.5cm, minimum height=0.56cm, inner sep=0pt,
        font=\sffamily\fontsize{6.8}{7.8}\selectfont},
    val/.style={circle, line width=0.6pt, minimum size=0.5cm, inner sep=0pt,
        font=\fontsize{7}{8}\selectfont},
    cflow/.style={->, line width=0.7pt, draw=cBlue},
    rflow/.style={->, line width=0.7pt, draw=cGreyM},
    lab/.style={font=\sffamily\fontsize{7}{8.2}\selectfont, text=cGreyD},
    row/.style={font=\sffamily\bfseries\fontsize{7.4}{8.8}\selectfont, anchor=west, inner sep=0pt, align=left},
    sub/.style={font=\sffamily\fontsize{7}{8.2}\selectfont, anchor=west, inner sep=0pt, align=left, text=cGreyD},
  ]

  \fill[cBlueL!38!white, rounded corners=3pt] (0,0.92) rectangle (12.72,2.64);   
  \fill[cBlueL!38!white, rounded corners=3pt] (0,3.76) rectangle (12.72,5.48);   

  \node[row, text=cGreyD] at (0.16,0.46) {Latent noise};
  \node[sub] at (0.16,0.12) {5 i.i.d.\ normals per step};
  \node[row, text=cBlue] at (0.16,2.04) {Generator $G$};
  \node[sub, text=cBlue] at (0.16,1.52) {LSTM with $u$ units,\\linear output and tanh};
  \node[row, text=cGreyD] at (0.16,3.37) {Paths};
  \node[sub] at (0.16,3.03) {generated or real};
  \node[row, text=cBlue] at (0.16,4.88) {Discriminator $D$};
  \node[sub, text=cBlue] at (0.16,4.36) {bidirectional LSTM with $u$\\units per direction, sigmoid};
  \node[row, text=cGreyD] at (0.16,6.21) {Per-step output};
  \node[sub] at (0.16,5.87) {probability of ``real''};

  \foreach \x/\t in {4.1/1, 6.5/2, 11.3/T} {
    \node[val, draw=cGreyM, fill=white] (z\t) at (\x,0.4) {$z_{\t}$};
    \node[gcell] (g\t) at (\x,1.46) {LSTM};
    \node[head] (h\t) at (\x,2.2) {tanh};
    \node[val, draw=cBlue, fill=cBlueL] (yh\t) at (\x-0.36,3.2) {$\hat{y}_{\t}$};
    \node[val, draw=cGreyM, fill=cGreyL] (y\t) at (\x+0.36,3.2) {$y_{\t}$};
    \draw[cflow] (z\t) -- (g\t);
    \draw[cflow, line width=0.6pt] (g\t) -- (h\t);
    \draw[cflow] (\x-0.36,2.38) -- (yh\t.south);
    \node[dcell] (c\t) at (\x,4.3) {BiLSTM};
    \node[head, fill=white] (s\t) at (\x,5.04) {sigmoid};
    \node[val, draw=cBlue, fill=white] (d\t) at (\x,6.04) {$d_{\t}$};
    \draw[cflow] (yh\t.north) -- (\x-0.36,4.02);
    \draw[rflow] (y\t.north) -- (\x+0.36,4.02);
    \draw[cflow, line width=0.6pt] (c\t) -- (s\t);
    \draw[cflow] (s\t.north) -- (d\t.south);
  }
  \foreach \y in {0.4, 1.46, 2.2, 3.2, 4.3, 5.04, 6.04} {
    \node[text=cBlue2, inner sep=0pt, minimum width=1.5cm] at (8.9,\y) {$\cdots$};
  }
  \draw[cflow] (g1) -- (g2);
  \draw[cflow] (g2.east) -- (8.2,1.46);
  \draw[cflow] (9.6,1.46) -- (gT.west);
  \draw[cflow, <->] (c1) -- (c2);
  \draw[cflow, <->] (c2.east) -- (8.2,4.3);
  \draw[cflow, <->] (9.6,4.3) -- (cT.west);
  \node[lab, anchor=south, inner sep=1.5pt, text=cBlue] at (5.3,1.46) {state};
  \node[lab, anchor=south, inner sep=1.5pt, text=cBlue] at (5.3,4.3) {states};

  \draw[cBlue, line width=0.6pt] (12.2,5.74) -- (12.36,5.74) -- (12.36,6.34) -- (12.2,6.34);
  \draw[cflow] (12.36,6.04) -- (13.02,6.04);
  \draw[draw=cGrey, fill=white, rounded corners=3pt, line width=0.6pt] (13.1,3.76) rectangle (16.5,6.5);
  \node[row, text=cGreyD] at (13.26,6.2) {Loss};
  \node[sub, text=cInk] at (13.26,5.62) {binary cross-entropy of\\$d_1,\dots,d_T$, averaged\\over the $T$ steps};
  \node[row, text=cGreyD] at (13.26,4.86) {Targets};
  \node[val, draw=cGreyM, fill=cGreyL, minimum size=0.3cm] at (13.42,4.44) {};
  \node[sub, text=cInk] at (13.7,4.44) {real path: 1};
  \node[val, draw=cBlue, fill=cBlueL, minimum size=0.3cm] at (13.42,4.02) {};
  \node[sub, text=cInk] at (13.7,4.02) {generated: 0 for $D$, 1 for $G$};

  \draw[draw=cGrey, fill=white, rounded corners=3pt, line width=0.6pt] (13.1,0.0) rectangle (16.5,3.44);
  \node[row, text=cGreyD] at (13.26,3.14) {Sampling};
  \node[sub, text=cInk] at (13.26,2.34) {after training, only $G$ is\\kept: a path $\hat{y}_1,\dots,\hat{y}_L$ of\\any length $L$ is obtained by\\unrolling the LSTM};
  \node[row, text=cGreyD] at (13.26,1.34) {Range};
  \node[sub, text=cInk] at (13.26,0.72) {the tanh output lies in $(-1,1)$;\\prices are mapped to this\\interval before training};
\end{tikzpicture}%